\documentclass[aps,prd,twocolumn,showpacs,groupedaddress]{revtex4}

\usepackage{graphicx}
\usepackage{dcolumn}
\usepackage{bm}

\usepackage{graphicx}
\usepackage{color}
\usepackage{slashed}
\usepackage{float}
\usepackage{amsmath}
\usepackage{bm}
\usepackage{amssymb}
\usepackage{multirow}
\usepackage{blindtext}
\usepackage{caption}
\usepackage{array}
\usepackage{siunitx}
\usepackage{breqn}

\def\apj #1 #2 #3 {#1, ApJ, {\bf #2}, #3}
\def\apjl #1 #2 #3 {#1, ApJ, {\bf #2}, L#3}
\def\apjs #1 #2 #3 {#1, ApJS, {\bf #2}, #3}
\def\aap #1 #2 #3 {#1, A\&A, {\bf #2}, #3}
\def\mnras #1 #2 #3 {#1, MNRAS, {\bf #2}, #3}
\def\pra #1 #2 #3 {#1, Phys.~Rev.~A., {\bf #2}, #3}
\def\prb #1 #2 #3 {#1, Phys.~Rev.~B., {\bf #2}, #3}
\def\prc #1 #2 #3 {#1, Phys.~Rev.~C., {\bf #2}, #3}
\def\prd #1 #2 #3 {#1, Phys.~Rev.~D., {\bf #2}, #3}
\def\pre #1 #2 #3 {#1, Phys.~Rev.~E., {\bf #2}, #3}
\def\prl #1 #2 #3 {#1, Phys.~Rev.~Lett., {\bf #2}, #3}
\def\plb #1 #2 #3 {#1, Phys.~Lett.~B., {\bf #2}, #3}
\def\science #1 #2 #3 {#1, Science., {\bf #2}, #3}
\def\nature #1 #2 #3 {#1, Nature., {\bf #2}, #3}
\def\nphysa #1 #2 #3 {#1, Nucl.~Phys.~A., {\bf #2}, #3}
\def\nphysb #1 #2 #3 {#1, Nucl.~Phys.~B., {\bf #2}, #3}
\def\nphysbs #1 #2 #3 {#1, Nucl.~Phys.~B.~Suppl., {\bf #2}, #3}

\def\h#1{\hbox{${}^{#1}$H}}

\def\h502{\hbox{$ h^{2}_{50}$}}

\def\fun#1#2{\lower3.6pt\vbox{\baselineskip0pt\lineskip.9pt
\ialign{$\mathsurround=0pt#1\hfil##\hfil$\crcr#2\crcr\sim\crcr}}}
\begin{document}

\title{Analysis of the Multi-component Relativistic Boltzmann equation for Electron scattering in Big Bang Nucleosynthesis}

\author{Nishanth Sasankan}
\email[]{nisaxaxa@gmail.com}
\affiliation{Department of Physics$,$ University of Notre Dame$,$ Notre Dame$,$ Indiana 46556}

\author{Atul Kedia}
\email[]{akedia@nd.edu, atulkedia93@gmail.com}
\affiliation{Department of Physics$,$ University of Notre Dame$,$ Notre Dame$,$ Indiana 46556}

\author{Motohiko Kusakabe}
\email[]{kusakabe@buaa.edu.cn}
\affiliation{ IRCBBC$,$ School of Physics$,$ Beihang University$,$ Beijing 100083 China}

\author{Grant J Mathews}
\email[]{gmathews@nd.edu}
\affiliation{Department of Physics$,$ University of Notre Dame$,$ Notre Dame$,$ Indiana 46556}


\date{\today} 

\begin{abstract}
Big-bang nucleosynthesis (BBN) is valuable as a means to constrain the physics of the early universe and it is the only probe of the radiation-dominated epoch. A fundamental assumption in BBN is that the nuclear velocity distributions obey Maxwell-Boltzmann (MB) statistics as they do in stars. Specifically, the BBN epoch is characterized by a dilute baryon plasma for which the velocity distribution of nuclei is mainly determined by the dominant Coulomb elastic scattering with mildly relativistic electrons. One must therefore deduce the momentum distribution for reacting nuclei from the multi-component relativistic Boltzmann equation. However, the full multi-component relativistic Boltzmann equation has only recently been analyzed and its solution has only been worked out in special cases. Moreover, a variety of schemes have been proposed that introduce non-thermal components into the BBN environment which can alter the thermal distribution of reacting nuclei. Here, we construct the relativistic Boltzmann equation in the context of BBN. We also derive a Langevin model and perform relativistic Monte-Carlo simulations which clarify the baryon distribution during BBN and can be used to analyze any relaxation from a non-thermal injection. We show by these analyses that the thermalization process leads to a nuclear distribution function that remains very close to MB statistics even during the most relativistic environment relevant to BBN. Hence, the predictions of standard BBN remain unchanged.
 \end{abstract}

\maketitle


\section{ Introduction}
Big-bang nucleosynthesis (BBN) remains as a pillar of modern cosmology\cite{bbnreview,Mathews17}. It provides an almost parameter free prediction of the abundances of the light isotopes $^2$H, $^3$He, $^4$He, and $^7$Li formed during the first few moments of cosmic expansion. At the onset of BBN ($T \sim 10^{10}$ K) the universe is mainly comprised of electrons, positrons, photons, neutrinos, and trace amounts of protons and neutrons. Once the temperature becomes low enough ($T \sim 10^9$ K) for the formation of deuterium, most neutrons are quickly absorbed by nuclear reactions to form $^4$He nuclei along with trace amounts of $^2$H, $^3$H, $^3$He, $^7$Li and $^7$Be. These trace amounts, however, are sensitive to the detailed freeze-out of the thermonuclear reaction rates as the universe cools. In this paper we re-examine the fundamental assumptions about the BBN epoch. In particular, we analyze the multi-component relativistic thermodynamics of the BBN environment.

Although the thermodynamics of both relativistic and nonrelativistic single-component gases have been known for many decades \cite{Juttner28}, the solution of the relativistic multi-component Boltzmann equation has only recently been attempted \cite{Kremer12,Kremer13} and transport coefficients have only been deduced for the case of equal or nearly identical-mass particles. An exact relativistic simulation has only been performed in one dimension to obtain the thermal equilibrium distribution functions of a two-component gas \cite{Cubero}. In three dimensions only a Fokker-Planck approximation for a Brownian particle in a relativistic bath has been developed to obtain the equilibrium distributions \cite{1+3}. 

Moreover, there has been recent interest in the possibility of a modification of the baryon distribution function from Maxwell Boltzmann (MB) statistics, in the form of Tsallis statistics \cite{Tsallis, Kusakabe19,Bertulani13, Hou17}, the influence of inhomogeneous primordial magnetic fields on baryons \cite{Luo19}, non-ideal plasma effects at low temperature \cite{Jang18}, the injection of nonthermal particles (e.g. \cite{Jedamzik04,Kawasaki05,Jedamzik06, Kusakabe09, Cyburt10, Voronchev12, Kusakabe14} and Refs.~therein), and small relativistic corrections to the MB distribution that arise due to nuclear kinetic drag \cite{McDermott18}. In the work of Ref.~\cite{McDermott18}, for example, the starting point was the Fermi-Dirac (FD) distribution for baryons from which corrections were deduced. Thus, it remains worthwhile to understand the evolution to thermalization of the relativistic multicomponent Boltzmann equation in the BBN environment.

The point of the present work, therefore, is to analyze the solution to the relativistic Boltzmann equation without an {\it a prior} assumption of what the baryon distribution should be. We show that the problem can be approximated as an ideal two component system of baryons immersed in a bath of relativistic electrons, for which the collision term is completely dominated by elastic scattering from relativistic electrons. We show that the thermalized baryon distribution is indeed close to MB statistics independently of the electron distribution function. This is verified by numerical Monte-Carlo simulations \cite{Kedia20} that can be used to follow the thermalization of the BBN environment. These simulations highlight the importance of correcting for the instantaneous viscosity experienced by recoiling nuclei. We also show that the assumption of kinetic equipartition (though relevant in the classical Langevin approximation, e.g.~\cite{Debbasch97,dunkel,Acosta}) is inappropriate for the relativistic primordial plasma.

\section{Big Bang Environment}
\subsection{Nuclear reaction rates}
The reaction rate between two species 1 and 2 can be written as \cite{Wagoner,Illiadis}
\begin{equation}
    R = n_1 n_2 \langle \sigma(v)v\rangle = n_1 n_2\int v\sigma(v) f(v) dv ~~,
\label{eq:1}
\end{equation}
where $n_1$ and $n_2$ are the number densities of the two species, $\sigma (v)$ is the reaction cross section, $v$ is the relative center-of-mass (CM) velocity and $f(v)$ is the relative velocity distribution function. In this paper we analyze the possible modification of $f(v)$ due to the unique environment encountered during BBN. Indeed, there has been considerable recent interest in deviations of the nuclear velocity distribution as a possible solution to the overproduction of lithium \cite{Kusakabe19, Bertulani13,Hou17}.


\subsection{Scattering in the background plasma}
At the start of BBN baryons are extremely dilute in number density compared to the background of $e^+-e^-$ pairs and photons. The baryon-to-photon ratio ($\eta$) is $\sim 10^{-9}$. Similarly, the ratio of baryons to $e^+-e^-$ pairs is $ \sim 10^{-9}$ during much of BBN. Hence, each nucleus undergoes scattering with a background plasma comprised of electrons, positrons and photons much more often than with other nuclei. This could be important when considering the relative velocity distribution functions $f(v)$ for nuclear reactions. That is, the velocity distributions of nuclei will result from scattering events with the mildly relativistic background plasma \cite{dunkel,Cubero} rather than with each other.

To justify the above statement regarding the relative scattering rates we first presume the usual thermodynamic relations for photons, electrons and nuclei during BBN. (This assumption will be revisited in Secs. \ref{secBoltz}, \ref{MC}, and the Appendix.) The number density of background photons is thus taken to be the usual Planck distribution:
\begin{equation}
    n_\gamma = \frac{g_\gamma}{2\pi^2\hbar^3 c^3}\int_0^{\infty} \frac{E^2}{e^{\frac{E}{kT}}-1}dE
   = \frac{2 \zeta(3) (kT)^3}{\pi^2\hbar^3c^3} ~~,
\end{equation}
where $c$ is the speed of light, $\hbar$ is the reduced Planck's constant, $k$ is the Boltzmann constant, $T$ is the temperature, $g_\gamma=2$ is the number of photon polarization states, $E$ is the photon energy.

Similarly, the number densities of positrons and electrons are described by a FD distribution,
\begin{equation}
    n_\pm = \frac{g_\pm}{\pi^2\hbar^3c^3}\int_0^{\infty} \frac{p^2}{\exp{\{(E \pm \mu)/kT\}}+1}dp ~~,
    \label{eq:e_dist}
\end{equation}
where $+~(-)$ denotes positrons (electrons), $g_\pm=2$ is the number of spin states, $E = \sqrt{p^2 + m_e^2}$ is the total energy with $m_e$ the electron rest mass, $p$ is the three momentum, and $\mu$ is the chemical potential for electrons. During most of BBN the chemical potential is small \cite{Wagoner}.

The elastic scattering cross section for photons with nuclei (Compton scattering using the Klein-Nishina formula) is given by
\begin{equation}
    \frac{d\sigma}{d\cos \theta} = \frac{\pi Z^4 \alpha^2}{(m c^2)^2} \left(\frac{\omega'}{\omega}\right)^2\left[\frac{\omega'}{\omega}+\frac{\omega}{\omega'}-\sin^2\theta\right] ~~,
\label{Compton}
\end{equation}
where, $\theta$ is the scattering angle, $\alpha$ is the fine structure constant, $Z$ is the nuclear charge, $m$ is the nuclear mass, $\omega$ and $\omega'$ are the frequencies of the incoming and outgoing photons, respectively.
From the angular integration of Eq.~(\ref{Compton}), the total reaction cross-section for a photon is $\sigma \le 66.5 \si{~fm^2} Z^4 (m_e/m)^2$.

The elastic scattering cross-section for electrons and positrons with nuclei is given by the Mott formula
\begin{equation}
    \frac{d\sigma}{d\cos\theta} = \frac{\pi Z^2 \alpha^2}{2 v^2p^2 \sin^4 \frac{\theta}{2}} \left(1 - \frac{v^2}{c^2} \sin^2 \frac{\theta}{2} \right) ~~,
\label{Mott}
\end{equation}
where $v$ is the velocity of the $e^-$ or $e^+$ particle.

The Coulomb scattering cross-sections can be evaluated using the Mott-formula or Rutherford-formula and is known to be infinite. However, a reasonable cut-off in the impact parameter for the incoming plasma particle is given by the Debye screening length $r_D = \sqrt{kT/4\pi n_0e^2}$ \cite{Jackson}. We adopt this as the maximum impact parameter to calculate the minimum scattering angle. Using these, we obtain two realistic approximations to the Coulomb cross sections: One is simply given by the area of a disk with radius $r_D$; while the second is based upon the Mott-formula with the upper limit defined by the minimum scattering angle.

Columns in Table \ref{table:rxn_rate} show the temperature dependence, respectively, for the ratio of number densities of electrons to photons $n_\pm/n_\gamma$, the electron-to-photon elastic-scattering cross-section ratio $\sigma_\pm/\sigma_\gamma$ for protons with cut-off radii at the Debye radius or the Mott formula minimum scattering angle, the ratio of nuclear scattering rates for electrons to photons $\Gamma_\pm/\Gamma_\gamma \equiv {n_\pm\sigma_\pm v_\pm}/{n_\gamma\sigma_\gamma c}$, and the ratio of rates for proton elastic scattering from electrons to elastic scattering from other protons, $\Gamma_\pm/\Gamma_p \equiv {n_\pm\sigma_\pm v_\pm}/{n_p\sigma_p v_p}$. It is evident from these ratios that nuclei scatter with the background $e^--e^+$ pair plasma significantly more than with photons or other nuclei during BBN. Hence, nuclei are overwhelmingly thermalized by elastic scattering with the background $e^--e^+$ pair plasma, while photons and other nuclei have a negligible effect on the thermodynamics.

\begin{table}
    \centering
    \caption{ Temperature dependence of various ratios relevant to proton elastic-scattering reaction rates with $e^--e^+$ plasma, photons and other protons. We use the minimum among the two cross section ratios ($4^{th}$ or $5^{th}$ column) to obtain the reaction rates for $e^--e^+$ plasma.}
    \begin{tabular}{| c | c | c | c | c | c | c |}
        \hline
        \multicolumn{2}{|c|}{$T$} & ${n_\pm}/{n_\gamma}$ & \multicolumn{2}{c|}{${\sigma_\pm}/{\sigma_\gamma}$} & ${\Gamma_\pm}/{\Gamma_\gamma}$ & ${\Gamma_\pm}/{\Gamma_p}$\\[0.5ex]
        \cline{1-2} \cline{4-5}
        $T_9$ & MeV & & $\sigma_\pm = \pi r_D^2$ & $\sigma_\pm = \sigma_{Mott}$ & & \\[1ex] \hline
        11.6 & 1 & 1.43 & $5\times10^4$ & $10^5$ & $10^5$ & $10^9$\\[1ex] \hline
        1.16 & 0.1 & $0.102 $ & $10^7$ & $10^5$ & $10^3$ & $10^{10}$\\[1ex] \hline
        0.116 & 0.01 &$10^{-13}$ & $2\times10^{28}$ & $10^{29}$ & $10^{14}$ & 10\\[1ex] \hline
    \end{tabular}
    \label{table:rxn_rate}
\end{table}

In what follows we model the response of nuclei to the dominant scattering from relativistic electrons via the multi-component relativistic Boltzmann equation. We also apply a Monte-Carlo simulation based upon the above scattering rates. In the appendix, we give a similar Langevin derivation. Indeed, the scattering rates in Table \ref{table:rxn_rate} suggest that the physical environment for BBN is similar to that of Brownian motion.

\section{Relativistic Boltzmann Equation}
\label{secBoltz}

For our purposes we can ignore the small corrections due to the cosmic expansion \cite{McDermott18}, and treat the space as flat. Following \cite{Kremer13} let us begin with a completely general mixture of $r$ constituents in a locally Minkowski space with metric tensor $\eta_{\alpha \beta} = diag(-1,1,1,1)$. The fluid consists of multiple particles of mass $m_a$ with $a = 1, ....r$. Each particle is characterized by space-time coordinates $x^\alpha$, $\alpha = 0,1,2,3$ and momenta $ p^\alpha_a = (E_a, p^{i}_a)$, so that $E_a = \gamma m_a = \sqrt{(p^i)^2 + m_a^2 }$ (we adopt natural units with $c = 1$). If we restrict our consideration to only elastic collisions, then the conservation of four momenta can be imposed
\begin{equation}
p^\alpha_a + p^\alpha_b = p^{'\alpha}_a + p^{'\alpha}_b ~~.
\end{equation}

The state of the mixture of $r$ relativistic species can be characterized by a set of one-particle distribution functions:
\begin{equation}
f({\bf x, p_a,} t) \equiv f_a~~, ~~~~~~a = 1,2,....r ~~.
\end{equation}

The total energy momentum tensor for the mixture is given by the sum of that due to each species
\begin{equation}
T^{\mu \nu} = \sum_{a=1}^r {T^{\mu \nu}_a}~~,
\end{equation}
where the contribution from each species is given in terms of the one particle distribution functions as:
\begin{equation}
T^{\mu \nu}_a =\int \frac{p_a^\mu p_a^\nu f_a}{p_a^0} d^3 p_a ~~.
\end{equation}

In the absence of external forces the one-particle distribution function characterizing collisions of constituent $a$ with constituent $b$ satisfies a Boltzmann equation,
\begin{eqnarray}
p^\alpha_a \partial_\alpha f_a &=& \sum_{b = 1}^r \int{\biggl[f_a' f_b' \biggl( 1 + \epsilon\frac{f_a h^3}{g_s}\biggr)\biggl( 1 + \epsilon\frac{f_b h^3}{g_s}\biggr) }\nonumber \\
&-&{ f_a f_b\biggl( 1 + \epsilon\frac{f_a' h^3}{g_s}\biggr)\biggl( 1 + \epsilon\frac{f_b' h^3}{g_s}\biggr)\biggr]} \\
&\times& {F_{ba} \sigma_{ab} d\Omega \frac{d^3p_b}{p_{b0}}} \nonumber~~,
\label{Boltz}
\end{eqnarray}
where the right-hand side is the one-particle collision term. The factors in parentheses account for the particle final state phase space with $\epsilon = +1$ for Bose-Einstein statistics and Pauli-blocking terms for $\epsilon = -1$ in Fermi-Dirac statistics. The quantity $h$ is the Planck constant. The quantity $g_s$ is the usual spin degeneracy factor appropriate to each species (not labeled here for simplicity). The quantity $F_{ba} =\sqrt{(p^\alpha_a p_{b \alpha})^2 - m_a m_b } $ is the invariant flux, while $\sigma_{ba}$ is the invariant differential elastic scattering cross section into an element of solid angle $ d\Omega$ that characterizes the collision of constituent $a$ with constituent $b$.

In a multi-component plasma, one must also count the flow of momentum and energy among components in the fluid. This leads to additional constraint equations of the moments of the distribution function \cite{Kremer13, Hebenstreit83}. However, as shown in Table \ref{table:rxn_rate} the collision term for nuclei is completely dominated by electron elastic scattering. Hence, one can reduce the multi-component relativistic Boltzmann equation to a two component system describing the scattering of relativistic electrons from nuclei. The identification of the thermodynamic variables can then be determined from the relativistic entropy flow as described below.

\subsection{General distribution function}
Denoting electrons $e$ and nuclei $n$, the relativistic baryon Boltzmann equation (\ref{Boltz}) becomes:
\begin{eqnarray}
\label{BoltzBe}
p^\alpha_n \partial_\alpha f_n &=& \int{\biggl[f_n' f_e' \biggl( 1 + \epsilon\frac{f_n h^3}{g_s}\biggr)\biggl( 1 + \epsilon\frac{f_e h^3}{g_s}\biggr) }\nonumber \\
&-&{ f_n f_e\biggl( 1 + \epsilon\frac{f_n' h^3}{g_s}\biggr)\biggl( 1 + \epsilon\frac{f_e' h^3}{g_s}\biggr)\biggr]} \\
&\times& {F_{e n} \sigma_{n e} d\Omega \frac{d^3p_e}{p_{e0}}} \nonumber~~,
\end{eqnarray}
Eq.~(\ref{BoltzBe}) differs from the usual one-particle Boltzmann equation in that the distribution is fixed by the dominant collisions with relativistic electrons. Nevertheless, from this one can immediately deduce the form of the stationary solution of the Boltzmann equation for the electrons and baryons.

For the distribution to be stationary one requires that the term in brackets on the right-hand side of Eq.~(\ref{BoltzBe}) vanish. Hence, 
\begin{eqnarray}
&&f_a' f_b' \biggl( 1 + \epsilon\frac{f_a h^3}{g_s}\biggr)\biggl( 1 + \epsilon\frac{f_b h^3}{g_s}\biggr)\nonumber \\
&=& { f_a f_b\biggl( 1 + \epsilon\frac{f_a' h^3}{g_s}\biggr)\biggl( 1 + \epsilon\frac{f_b' h^3}{g_s}\biggr)}~~.
\label{feq}
\end{eqnarray}
Then, taking the logarithm of both sides of Eq.~(\ref{feq}) one has for both the electrons and baryons the form $\ln {[f/(1 + \epsilon f h^3/g_s)]} = A - B _{ \alpha} p^\alpha$, which is a summational invariant \cite{Kremer13} for which the terms $A$ and $B_\alpha$ are determined from the stationary values of the particle four-flow and the energy-momentum tensor. The stationary distribution for both baryons and electrons is then of the form \cite{Kremer13} :
\begin{equation}
f({\bf p}) =\frac{g_s/h^3}{ \exp{[-a + B^\alpha p_\alpha]} - \epsilon}~~,
\end{equation}
where $a=A +\ln(h^3/g_s)$.

Next, consider the particle number-density four-current $J^\mu = n U^\mu$, with $n$ the local proper rest particle density and $U^\mu$ the particle four velocity, with $U_\mu U^\mu = -1$. Since $J^\mu$ is the only relevant four vector, one can identify $B^\mu \propto J^\mu \ = \zeta U^\mu$. Then the equilibrium distribution takes the form:
\begin{equation}
f_{\rm eq}({\bf p}) = \frac{g_s/h^3}{\exp{\bigl[-a + \zeta \bigl(U^\alpha p_\alpha \bigr) \bigr]} - \epsilon}~~.
\label{MJ}
\end{equation}
 
\subsection{Entropy flow and the Gibbs equation}
For the next step one must identify the relation between the parameter $\zeta$ and the temperature $T$. To do this one must define the thermodynamic variables via the Gibbs relation:
\begin{equation}
 ds_E = \frac{1}{T} \bigl( de - \frac{P}{n^2}dn\bigr) ~~,
 \label{dsGibbs}
\end{equation}
where $s_E$ is the equilibrium entropy per particle. The total internal energy per particle is $e = \langle E \rangle = \langle \gamma m \rangle$, $P$ is the pressure, and $n$ is the number density. The total equilibrium relativistic entropy is deduced from the entropy flow per particle $s_E$ as
\begin{equation}
S_E^\alpha = n s_E U^\alpha ~~.
\label{SE}
\end{equation}
The total entropy flow, however, must be determined from the general distribution function [Eq.~(\ref{MJ})] $f$ \cite{Kremer-book}.
\begin{eqnarray}
S_E^\alpha &=& -k \int p^\alpha f \biggl[ \ln{\biggl( \frac{f h^3}{ g_s}\biggr)}\nonumber \\
&- &\biggl( 1 + \frac{g_s}{\epsilon f_a h^3}\biggr)ln{\biggl( 1 + \epsilon\frac{f_a h^3}{g_s}\biggr)}\biggr] \frac{d^3p}{p_0}~~.
\label{Stot}\end{eqnarray}

Insertion of the distribution function Eq.~(\ref{MJ}) into Eqs.~(\ref{SE}) and (\ref{Stot}) leads to the following expression for the entropy per particle \cite{Kremer-book}.
\begin{equation}
s_E = k \biggl( \frac{ \zeta}{m} e - a + \frac{4}{3} \pi \frac{m^4}{nT} \frac{g_s}{h^3}J_{40} \biggr)~~,
\label{dsMJ}
\end{equation}
where, $a = \mu_E/kT $ is a constant of integration related to the chemical potential at equilibrium, and 
\begin{equation}
J_{mn}(\zeta, \mu_e/kT) = \int_0^\infty \frac{\sinh^n{\theta} \cosh^m{\theta}}{\exp{(-\mu_e/kT + \zeta \cosh{\theta})} - \epsilon} d \theta ~~.
\end{equation}

Following Ref.~\cite{Kremer-book}, we show below that Eq.~(\ref{dsMJ}) can be reduced to the classical Sackur-Tetrode equation in the non-degenerate non-relativistic limit. However, to solve for $\zeta$ we only need the differential form to compare with the Gibbs relation, Eq.~(\ref{dsGibbs}):
\begin{equation}
  ds_E = \frac{k \zeta}{m} \biggl( de - \frac{P}{n^2}dn\biggr)
 \label{dszeta}
\end{equation}

We proceed to show below via analytic and numerical simulations that for any two-component system in temperature equilibrium, one can identify Eqs. (\ref{dsGibbs}) and (\ref{dszeta}) so that $\zeta = m/T$ even if one component is relativistic and one component is nonrelativistic. However, one can imagine stationary situations in which the identification $\zeta = m/T$ is not possible. This could happen, for example, via the continual injection of a non-thermal spectrum of particles that keeps one component of the system out of temperature equilibrium with the background thermal plasma \cite{Jedamzik04,Kawasaki05,Jedamzik06, Kusakabe09, Cyburt10, Voronchev12, Kusakabe14}.

Now, from the energy-momentum tensor relations
\begin{equation}
ne = T^{\mu \nu} U_\mu U_\nu ~~,
\end{equation}
and
\begin{equation}
-ne + 3p = T^{\mu \nu} \eta_{\alpha \beta} ~~,
\end{equation}
the relevant state variables are then:
\begin{equation}
n = 4 \pi m^3 \frac{g_s}{h^3} J_{21}(\zeta)~~~,
\end{equation}
\begin{equation}
e = m \biggl(\frac{J_{22}(\zeta)}{J_{21}(\zeta)} \biggr)~~,
\end{equation}
\begin{equation}
P= \frac{4}{3} \pi m^4 \frac{g_s}{h^3}J_{40}(\zeta) ~~.
\end{equation}

Note, that it is not possible to obtain an explicit expression for the relation between $\zeta$ and temperature directly from the distribution function \cite{Kremer-book}. To obtain $\zeta$ one must consider the physics of the environment for each species of the multicomponent system.

\subsubsection{Relativistic non-degenerate electrons}

First we consider the electrons. Early during BBN the electrons interact much more frequently with each other than with nuclei. Thus, they can essentially be treated as a single component relativistic gas. In this limit one can simply equate Eqs.~(\ref{dsGibbs}) and (\ref{dszeta}) so that $\zeta_e = m_e/kT$. In the cosmological rest frame $U^\alpha p_\alpha = E_e/m_e$ is the total relativistic electron energy. Hence, the Fermi-Dirac distribution for the electrons is obtained.
\begin{equation}
f_e(E) = \frac{g_s/h^3}{\exp( (-\mu_e + E_e)/kT) + 1}~~.
\end{equation}
 However, the FD distribution is notoriously difficult to integrate to obtain the thermodynamic variables. Nevertheless, the nondegenerate limit, $(E_e - \mu_e)/kT \gg 1$, is appropriate for the reaction rates of big bang nucleosynthesis. For a non-degenerate gas, the $J_{m n}$ can be related \cite{Kremer-book} to modified Bessel functions of the second kind $K_n$. In this case the electrons can be represented by a Maxwell-J\"uttner distribution \cite{Juttner28} and the thermodynamic variables can be reduced to \cite{Kremer-book}:
\begin{equation}
 n_e = 4 \pi m_e^2 kT \frac{g_s}{h^3} K_2(m_e/kT)e^{\mu_e/kT}~~~,
\end{equation}
\begin{equation}
e_e = m_e \biggl(\frac{K_3(m_e/kT)}{K_2(m_e/kT)} - \frac{kT}{m_e} \biggr)~~,
\label{ee:eq}
\end{equation}
\begin{equation}
P_e = 4 \pi m_e^2 (kT)^2 \frac{g_s}{h^3}K_2(m_e/kT)e^{\mu_e/kT} = n_e kT~.
\label{pe}
\end{equation}
We note that Eq.~(\ref{pe}) is not true for a relativistic FD gas, but only holds in the non-degenerate limit appropriate here.

\subsubsection{Nuclei experiencing elastic collisions with electrons}

The physics of the nuclei, however, is different in this scenario. The isotropic velocities of the nuclear component of the cosmic fluid during BBN are dominated (at least initially) by collisions with relativistic electrons rather than other baryons. To deduce the baryonic pressure one must consider that elastic scattering with electrons conserves momentum and energy.

The derivation of the pressure for the nuclei is straightforward. For a system of discrete point particles, the energy-momentum tensor takes the form
 \begin{equation}
 T^{\mu \nu} = \sum_a \frac{p^\mu(a) p^\nu(a)}{p^0(a)} \delta^{(3)}(\vec x - \vec x(a)) ~~,
 \end{equation}
where now $a$ labels each particle and $p^\mu (a)= m_a U^\mu(a) $ is the four momentum, and in flat space $U^\mu = (\gamma, \gamma v^1, \gamma v^2, \gamma v^3)$.
 
One is only interested in the spatial components $T^{i j}$ for the derivation of pressure in the cosmological rest frame. Moreover, since the spatial components of momentum are isotropic, only diagonal components are relevant. Hence we can write
 \begin{eqnarray}
 P_n = T_n^{i i} &=& \sum_a \frac{p^i(a) p^i(a)}{p^0(a)} \delta^{(3)}(\vec x - \vec x(a))\nonumber \\
 & = & \sum_a \gamma_a m_n (v^i_a)^2 \delta^{(3)}(\vec x - \vec x(a))~~\nonumber \\
 &= &\frac{1}{3} n_n m_n \langle \gamma v^2 \rangle ~~,
 \label{pb}
 \end{eqnarray}
where the factor of 1/3 follows from the isotropy of the frame at rest w.r.t. the cosmic fluid,
\begin{equation}
\langle \gamma v_x^2 \rangle =\langle \gamma v_y^2\rangle = \langle \gamma v_z^2\rangle = \frac{1}{3} \langle \gamma v^2 \rangle~~.
\label{pressure_jutt}
\end{equation}
A similar derivation applies to electrons, i.e.
\begin{equation}
 P_e = T^{i i}_e = \frac{1}{3} n_e m_e \langle \gamma v^2 \rangle~~.
 \end{equation}

However, in thermal equilibrium baryons and electrons, are at the same temperature. Moreover, the average $ \langle \gamma v^2 \rangle$ for each species must be the same. So from Eq.~(\ref{pe}), the pressure per baryon is
\begin{equation}
\frac{P_n}{n_n} = \frac{P_e }{n_e} = kT ~~.
\label{pbn}
\end{equation}

Indeed, using the non-degenerate distribution to evaluate the average in Eq.(\ref{pb}) identically gives
\begin{equation}
P_n = n_nkT ~~.
\end{equation}

\subsubsection{ Internal energy of the nuclear fluid}

Having derived the pressure, the average energy per nucleus is almost trivial.
\begin{equation}
e_n = \gamma m_n = m_n + (\gamma-1)m_n \approx m_n + (1/2) m_n \langle v_n^2 \rangle ~~,
\label{eb}
\end{equation}
where the latter approximation follows from the fact that in the BBN epoch, $v_n << 1$.

From Eqs.~(\ref{pb}), (\ref{pbn}), and (\ref{eb}) it follows that 
\begin{equation}
e_n = m_n + \frac{3}{2} kT~~.
\label{eneq}
\end{equation}
Hence, even in this idealized case of nuclei only experiencing elastic scattering from a distribution of relativistic electrons, the baryons have the same average kinetic energy as that of a classical Maxwell-Boltzmann gas for which $\langle m_n v_n^2/2 \rangle = (3/2)kT$. Indeed, this result is independent of the electron distribution function as long as the baryons are in temperature equilibrium with electrons.

The Gibbs relation for nuclei in the relativistic electron bath is then satisfied with $\zeta_n = m_n/kT$. Then in the non-degenerate limit, the equilibrium chemical potential becomes
\begin{equation}
\mu_E = kT \ln{\biggl[\frac{n h^3}{4 \pi g_s m_n^2 k T K_2(\zeta)}\biggr]} ~~,
\end{equation}
so that in the non-relativistic limit ($\zeta >>1$) the entropy per particle reduces to the classical Sackur-Tetrode equation,
\begin{equation}
\frac{s_E}{k} = \biggl(\ln{\frac{T^{3/2}}{n}} - \ln {\biggl[\frac{h^3}{g_s (2 \pi m k)^{3/2}}\biggr]}+ \frac{5}{2 }\biggr)~~.
\end{equation}
Finally, for $v<<1$, $U^\alpha \approx (1,0,0,0)$, the non-degenerate distribution for nuclei reduces to the usual MB kinetic energy distribution,
\begin{equation}
 f_n(E) = \frac{n}{(2 \pi kT )^{3/2}}\exp{\left(-\frac{ m_n v^2}{2 kT}\right)}~~.
\label{MBf}
\end{equation}

Hence, in the limit of nuclei dominated by relativistic electron elastic scattering we have shown analytically that the standard MB statistics emerges. Note that this result is independent of the electron distribution function. The only requirement is thermal equilibrium with the electron gas. We note here, that a Langevin derivation assuming kinetic energy equipartition as is often done in classical analyses \cite{dunkel} cannot be applied for a background of relativistic particles. However, thermal equilibrium as employed here is more relevant for a relativistic plasma. On the other hand, this analytic derivation is in the non-degenerate limit so that use could be made of the analytic properties of the Maxwell-J\"uttner distribution. We have ignored the effect of quantum statistics. Therefore, we check this result with Monte-Carlo simulations utilizing both a FD and MJ distributions as described below in Sec. \ref{MC} and in Ref.~\cite{Kedia20}.

\section{Monte-Carlo Scattering Simulation}
\label{MC}
As a test of our statistical analysis we created a Monte-Carlo simulation of the Brownian motion of a proton during BBN \cite{Kedia20}. That is, we simulated nuclear thermalization in a bath with temperatures and an environment relevant to BBN. This was done to numerically obtain the true multi-component velocity distributions for nuclei. Table \ref{table:rxn_rate} showed that photons play a negligible role in this process. Hence, we only needed to simulate scattering of a relativistic FD distribution of $e^- - e^+$ pairs with nuclei. During this scattering process energy is transferred to or from nuclei. The direction of transfer of momentum is governed by the angle of incoming particles, the velocity of incoming particles and the scattering angle of the outgoing electron or positron. For our simulation the angle of the incoming particles was chosen isotropically in the cosmic frame. However, this would not in general be isotropic in the nuclear rest frame due to the accumulated nuclear recoil velocity.

We randomly selected the incoming electron momentum from the FD distribution. The angle of scattering for electrons was weighted by the differential cross-section in Eq.~(\ref{Mott}). The reactions were simulated in three dimensions. The incoming momentum of nuclei before each scattering event was given by its momentum after the previous scattering event. The scattering process was then repeated for a sufficiently large number of times ($\sim 10^7$). Note that according to Table \ref{table:rxn_rate} that even in the worst case at $kT = 0.1$ MeV there would only be $< 10^{-3}$ photon scatterings for each electron scattering. Moreover, for a baryon-to-photon ratio of $\eta \sim 10^{-9}$, there would be no nucleus-nucleus scatterings during 10$^7$ electron collisions. Hence, the influence of nuclear and photon scattering is negligible. This, however, is not the case in stars where the baryon density is much higher.

We note, as demonstrated in \cite{Kedia20}, that it is important to account for the effect of the instantaneous viscosity (i.e electrons moving opposite to the nuclear direction of motion collide more frequently with the nucleus). This was corrected by sampling the electrons from the electron flux at a rate proportional to $vf(v)$, where $v$ is the relative velocity in the frame of the nucleus [cf. Eq.~(\ref{eq:1})]. With this correction the nuclear distribution function is skewed to lower energies due to the increase in the collision rate along the direction of motion. This reduces the high-energy tail of the distribution such that the resultant distribution overlaps well with MB statistics rather than the electron FD distribution.

\begin{figure}
\includegraphics[height=2.3in,width=3.5in]{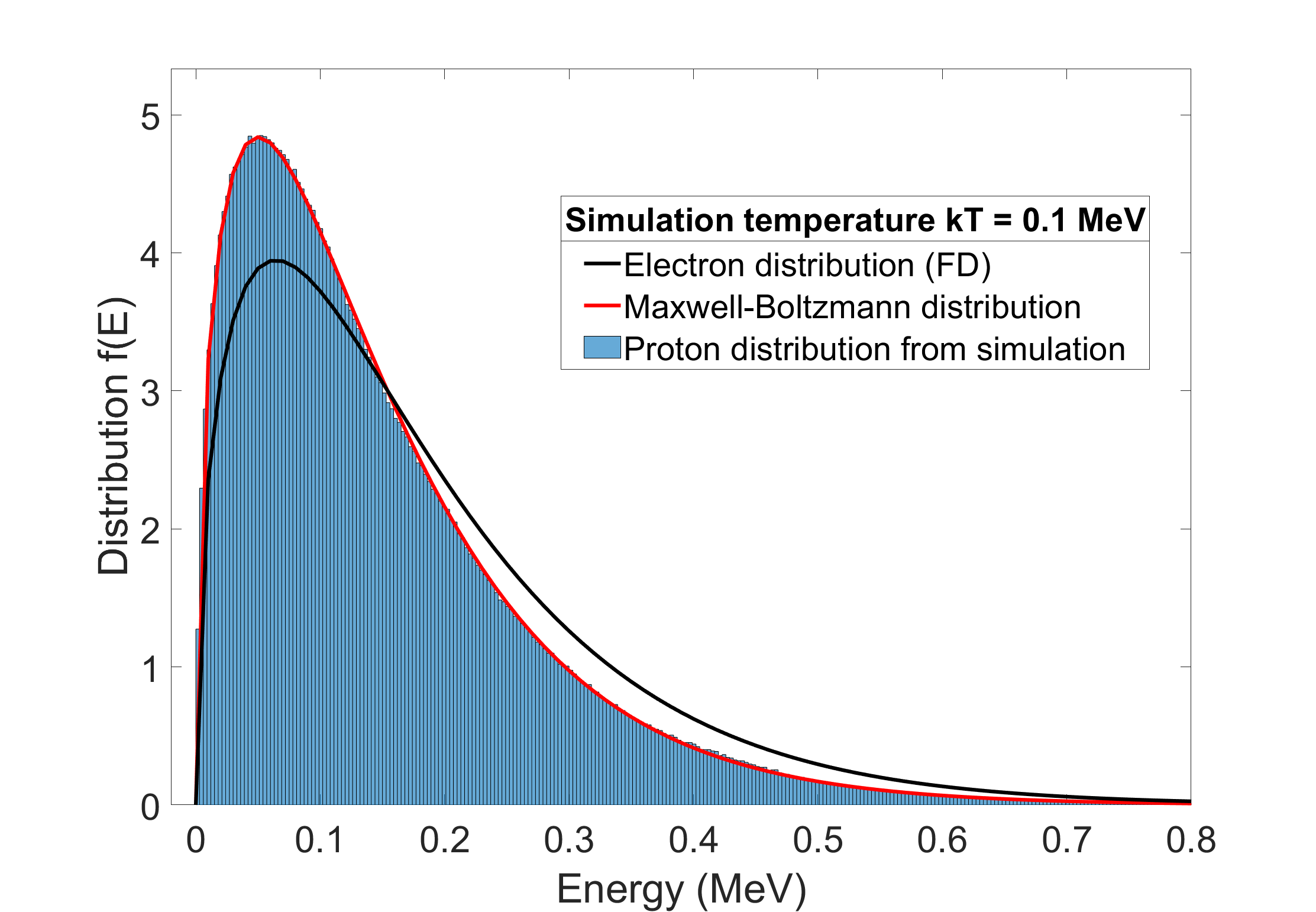}
\includegraphics[height=2.3in,width=3.5in]{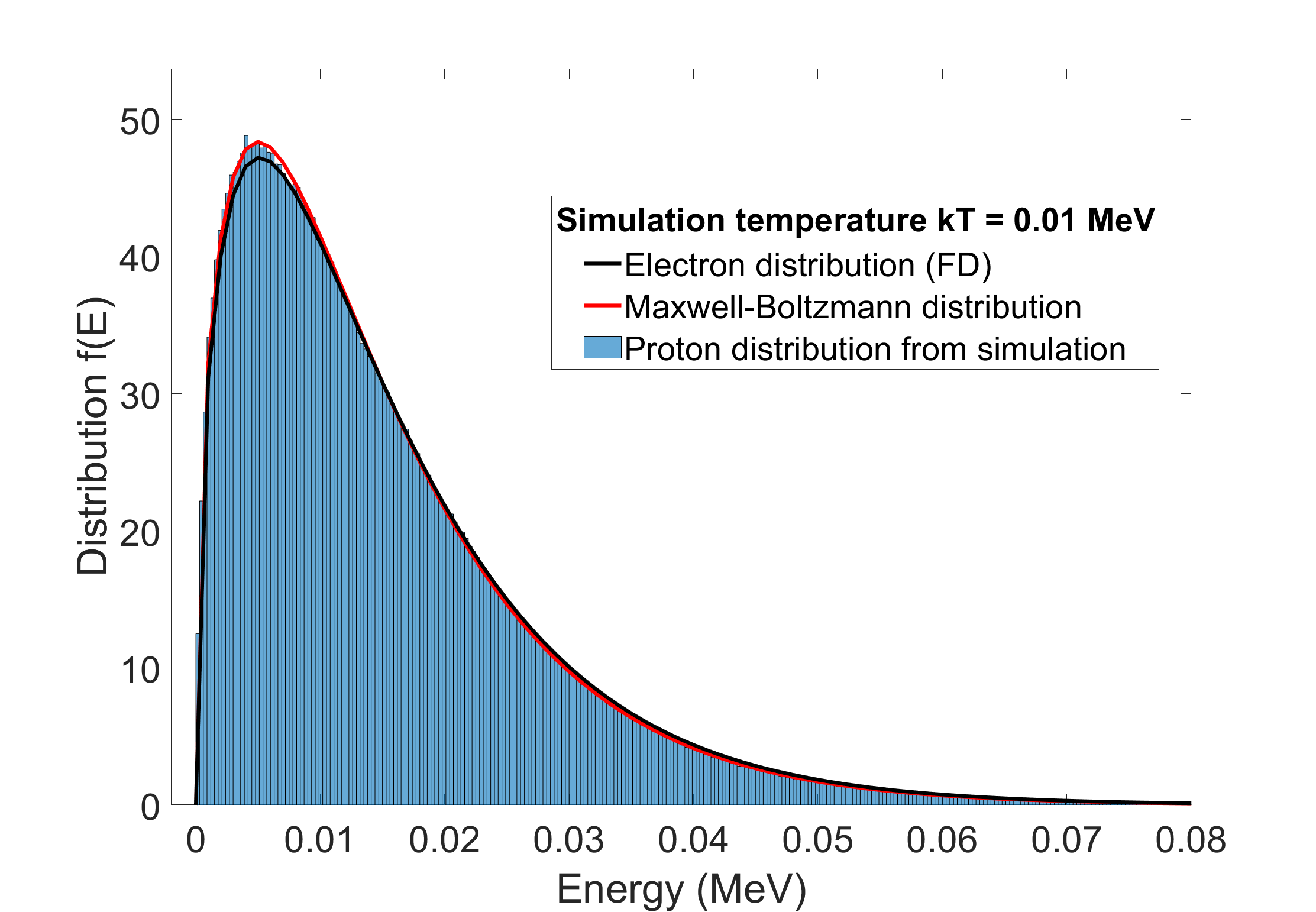}
\caption{Monte-Carlo histograms (blue bars) of the kinetic energy distribution of baryons scattering in a bath of mildly relativistic FD $e^+-e^-$ plasma (black line) at kT = 0.1 MeV (upper panel) and kT = 0.01 MeV (lower panel) compared to the kinetic energy distribution of a classical Maxwell-Boltzmann distribution (red line). [Color online]}
\label{fig:1}
\end{figure}

The upper panel in Fig. \ref{fig:1} shows a Monte-Carlo simulation \cite{Kedia20} of the kinetic energy distribution of protons in a bath of 0.1 MeV FD relativistic electrons after a large number of simulated elastic scattering events. This temperature roughly corresponds to the start of the BBN nuclear reaction epoch. The lower panel shows a similar result for $kT = 0.01$ MeV roughly corresponding to the end of the BBN epoch. Also shown for illustration is the distribution of a classical MB gas and the FD distribution of electrons. From this it is clear that even at the highest temperatures of the BBN epoch, in the idealized case of dilute charged baryons elastically scattering from relativistic electrons, the baryon distribution functions are very close to that of a classical Maxwell Boltzmann gas. Indeed, an analysis of the standard deviations of the proton distribution from a pure MB distribution is about one percent over the interval from $0.25kT$ to $3kT$ in energy. This is consistent with numerical and statistical fluctuations in the simulation. 

Using the same simulation technique we have checked \cite{Kedia20} that the thermalized nuclear distribution function is that of MB statistics independently of the electron distribution function. Even a delta-function electron distribution function will lead to an MB distribution for nuclei, consistent with our analysis of the relativistic Boltzmann equation.

\section{Discussion and Conclusions}

In summary, we have shown that the thermalization of nuclei during BBN is dominated by Coulomb elastic scattering with the background mildly relativistic $e^+-e^-$ pair plasma. Hence, even though there are photons and other nuclei present during the era, these don't contribute significantly toward the thermalization of the nuclear distribution functions. Moreover, we have shown from a solution to the Relativistic multi-component Boltzmann equation that the equilibrium distribution of nuclei in the $e^+-e^-$ pair plasma remains very close to MB statistics. The solution to the Boltzmann equation is confirmed via a Monte-Carlo thermalization simulation that also recovers a nuclear MB distribution function independently of the electron distribution function. An important reason for this result is the effect of the instantaneous viscosity due the motion of the baryons w.r.t. the background plasma as shown in \cite{Kedia20}.

For completeness, in the appendix we also discuss a Langevin Brownian-motion derivation with the imposition of thermal equilibrium rather than kinetic-energy equipartition. We show that this also leads to a MB distribution for nuclei in the $e^+-e^-$ pair plasma.

\section{Acknowledgments}
Work at the University of Notre Dame is supported by the U.S. Department of Energy under Nuclear Theory Grant DE-FG02-95-ER40934. One of the authors (M.K.) acknowledges support from the Japan Society for the Promotion of Science (27.957). The authors wish to thank an anonymous referee for pointing out the weakness of assuming kinetic equipartition and the need to account for the baryon recoil motion w.r.t. the background plasma.


\section{Appendix A}
For completeness of the theory of the thermalization of a multi-component relativistic gas we here derive a Langevin model for the distribution function for heavy nuclei in a bath of light relativistic electrons.

In one dimension the Langevin model for Brownian motion obeys the equation of motion
\begin{equation}
    m\dot{v}=-\lambda v+R(t)~~.
\label{Langevin1}
\end{equation}
Here, $m$ is the mass of the particle, $v$ is the velocity, $\lambda$ is a drag coefficient, and $R(t)$ is a noise term representing the effect of collisions with the background fluid at time $t$. The force $R(t)$ has a Gaussian probability distribution centered around $R=0$ and the value at time $t+\tau$ does not depend on the value at time $t$, i.e.
\begin{equation}
P(R) = \frac{1}{\sqrt{2 \pi \langle R(t)^2 \rangle}} \exp{\biggl[ \frac{-R^2}{2 \langle R(t)^2 \rangle} \biggr]} ~~,
\end{equation}
\begin{equation}
\langle R(t)\rangle = 0~~, 
\end{equation}
and
\begin{equation}
\langle R(t) R(t + \tau) \rangle = \langle R(t)^2 \rangle \delta(\tau) ~~.
\end{equation}
These conditions are easily satisfied in the BBN scattering environment. Note also, that it does not matter whether $R(t)$ is due to scattering from relativistic or non-relativistic particles as long as the force has a Gaussian probability distribution, the Langevin formalism can be applied to derive the distribution function of the massive particle. Indeed, massive particles in a relativistic fluid do experience a random Gaussian force as has been shown in Ref.~\cite{ Plyukhin}.

The general solution to Eq.~(\ref{Langevin1}) is given by
\begin{equation}
v(t)=v_0\exp\biggl(\frac{-\lambda t}{m}\biggr) + \frac{1}{m}\int_0^tR(t')\exp\biggl(\frac{-\lambda (t-t')}{m}\biggr)dt' ~~.
\label{veq}
\end{equation}
Even without specifying the explicit form of $R(t)$, one can deduce average properties of $v(t)$. In particular, from Eq.~(\ref{veq}) one can take the limit 
as $t\rightarrow\infty$, to obtain:
\begin{eqnarray}
 \langle v^{2}(t) \rangle =\frac{q}{2\lambda m} ~~,
\label{v2eq}
\end{eqnarray}
where $q = \langle R(t)^2 \rangle \delta(\tau) $ and $\langle R(t)^2 \rangle$ is the variance of $R(t)$.

Now, as shown in Eqs.~(\ref{pbn})-(\ref{eneq}) the temperature equilibrium between the non-relativistic baryons and relativistic background requires:
\begin{equation}
    \frac{1}{2}m_n \langle v^{2} \rangle =\frac{3}{2}kT~~.
\end{equation}

Then using Eq.~(\ref{v2eq}) one has,
\begin{equation}
    \frac{1}{2}m_n \langle v^{2} \rangle =\frac{q}{4\lambda}=\frac{3}{2}kT~~,
\end{equation}
so that
\begin{equation}
    q= 6 \lambda kT~~.
\end{equation}

The Langevin evolution of the velocity distribution function $f(v)$ reduces to a Fokker-Planck equation of the form
\begin{equation}
    \frac{\partial f(v,t)}{\partial t}=\lambda\frac{\partial (vf(v,t))}{\partial v} + \lambda \frac{kT}{m}\frac{\partial^{2}f(v,t)}{\partial v^{2}}~~.
\end{equation}
At equilibrium $\frac{\partial f(v,t)}{\partial t}=0$, so that
\begin{equation}
   \frac{\partial (vf(v,t))}{\partial v} + \frac{kT}{m}\frac{\partial^{2}f(v,t)}{\partial v^{2}}=0. 
\end{equation}
Notice that this is independent of the drag term $\lambda$. The solution for $f(v)$ then takes the form
\begin{equation}
    f(v) \propto \exp{\biggl(-\frac{mv^{2}}{2kT}\biggr)} ~~.
\end{equation}
On normalizing one obtains the usual MB distribution
\begin{eqnarray}
 f(v)&=&\left(\frac{m}{2\pi kT}\right)^{\frac{3}{2}}4\pi v^{2}\exp{ \left(-\frac{mv^{2}}{2 kT}\right)} ~~, \\
\label{fvmodeq}
    f(E)&=&2\left(\frac{1}{kT}\right)^{\frac{3}{2}}\sqrt{\frac{E}{\pi}}\exp{\left(-\frac{E}{kT}\right)} ~~ .
\label{femodeq}
\end{eqnarray}
Hence, for a nucleus in equilibrium with a relativistic background $e^+-e^-$ plasma, the distribution function can be described as the usual Maxwell-Boltzmann distribution.

\end{document}